\documentclass[%
 reprint,
 amsmath,amssymb,
 aps,
]{revtex4-2}

\usepackage{graphicx}
\usepackage{dcolumn}
\usepackage{bm}

\usepackage{lineno,hyperref}
\usepackage{braket}
\usepackage{amsmath,amssymb}
\usepackage{physics}
\usepackage[T1]{fontenc}

\begin{document}

\preprint{APS/123-QED}

\title{Simultaneous Suppression of Thermal Phase Noise and Relative Intensity Noise in a Fiber Optic Gyroscope}

\author{Nobuyuki Takei$^{1,2}$}
\author{Martin Miranda$^{1}$}
\author{Yuki Miyazawa$^{1}$}
\author{Mikio Kozuma$^{1,3}$}

\affiliation{%
	$^{1}$Institute of Innovative Research, Tokyo Institute of Technology, 4259 Nagatsuta, Midori, Yokohama, Kanagawa 226-8503, Japan}
	
\affiliation{%
	$^{2}$MIZUSAQI Inc., 1-48-3 Hanasaki, Naka, Yokohama, Kanagawa 231-0063, Japan}
	
\affiliation{%
	$^{3}$Department of Physics, Tokyo Institute of Technology, 2-12-1 O-okayama, Meguro, Tokyo 152-8550, Japan}


\begin{abstract}
The short-term sensitivity of a several-kilometers long fiber-optic gyroscope is limited mainly by thermal phase noise and relative intensity noise.
Increasing the phase modulation frequency decreases the thermal phase noise but not the relative intensity noise since it behaves as white noise.
Here, we propose and experimentally demonstrate that the angular random walk can be effectively decreased by suppressing relative intensity noise at the modulation frequency and its third-order harmonic using direct feedback to the drive current of a superluminescent diode.
Our simultaneous suppression of thermal phase noise and relative intensity noise yields an angular random walk of $15\,\mu\mathrm{deg}/\sqrt{\mathrm{h}}$ and a bias instability of $33\,\mu\mathrm{deg}/\mathrm{h}$ using a fiber coil with a length of $5$\,km and an effective area of $280\,\mathrm{m^2}$ for a measurement time of 40 hours.
\end{abstract}

\maketitle

\section{Introduction}
An interferometric fiber-optic gyroscope (FOG) is a device that measures the rotation rate by making use of the Sagnac effect, which induces a phase difference between two counterpropagating light waves traveling in the same fiber coil~\cite{Lefevre2014,Lefevre2020,Korkishko2017}.
Because of their high precision and stability, FOGs have been widely used in numerous applications, such as satellite stabilization, inertial navigation, and rotational seismometers~\cite{Lefevre2016,Sanders2016,Napoli2016}.
In evaluating the performance of FOGs, two aspects are often considered: long-term stability and short-term sensitivity.
Long-term stability is crucial, for example, in terms of reducing drift and errors that can accumulate when FOGs are operated in an actual inertial navigation system (INS)~\cite{Paturel2014}.
Better short-term sensitivity not only improves the performance of INSs but also expands the application of FOGs to other areas, such as seismology~\cite{Schmelzbach2018,Kurzych2018}.
One way to improve the short-term sensitivity is to increase the scaling factor by using a larger fiber coil~\cite{Toldi2017,Li2019}.
However, optical power is attenuated in such a long fiber, and thereby the effect of shot noise could degrade the sensitivity~\cite{Guattari2016}.
In addition, thermal phase noise becomes nonnegligible for long fibers~\cite{Li2019,Logozinskii1981,Knudsen1995,Moeller1996}.

Another method is to suppress the noises that impose a limitation on the short-term sensitivity~\cite{Morris2022}.
Four types of noise limit the sensitivity of a several-kilometers long FOG: thermal phase noise (TPN), relative intensity noise (RIN), shot noise, and detection noise.
Since the TPN originates from the thermodynamic fluctuation of the optical fiber, it decreases as the modulation frequency increases~\cite{Li2019,Knudsen1995,Moeller1996}.
The TPN was successfully reduced in an experiment using a FOG with a 30\,km-long single-mode (SM) fiber coil, where modulation frequency corresponding to the 33rd-order harmonic of the eigenfrequency was employed~\cite{Li2019}.
However, the RIN eventually limited the angular random walk in this experiment.
In modern FOGs, broadband light sources with low temporal coherence are the preferred choice since they reduce coherence-induced noise.
However, the random beating between frequency components within the broad spectrum causes significant intensity noise called RIN~\cite{Burns1990,Burns1996,Shin2010,Blake1994}.
In comparison to the shot noise level of the light source, this RIN is often referred to as excess RIN~\cite{Burns1990}, and many methods of reducing or compensating for the excess RIN have been proposed and tested~\cite{Shin2010,Blake1994,Yurek1986,Moeller1991,Guattari2014,Polynkin2000,Rabelo2000,Honthaas2014,He2020,Suo2021}.
One example is a technique called noise subtraction, which can be realized electronically~\cite{Yurek1986,Moeller1991} or optically~\cite{Guattari2014,Polynkin2000,Rabelo2000}.
In both cases, some portion of the light source is picked up as a reference, and its associated noise is considered to be correlated to that of the output power from the FOG.
By carefully adjusting the delay between the reference and the output, the noise in the output is compensated for with that in the reference.
Many other efforts for RIN suppression have been made, such as optical spectrum filtering~\cite{Honthaas2014}, a dual-polarization scheme~\cite{He2020}, and a technique employing a semiconductor optical amplifier~\cite{Shin2010,Suo2021}.
When the TPN and excess RIN are sufficiently suppressed, the shot noise of the light source and detection noise finally limit the sensitivity.

In this paper, we propose and experimentally demonstrate simultaneous suppression of TPN and RIN.
We reduce TPN by employing a relatively high phase modulation frequency. In the case of a square-wave modulation, undesired spikes appear in a FOG signal, resulting in additional noise. This noise becomes more prominent for higher-frequency phase modulation. Therefore, we use a sinusoidal-wave modulation scheme.

On the other hand, we suppress RIN by examining the property of RIN in a sinusoidal-wave modulation scheme. As is discussed in the next section, it is necessary to suppress RIN components at a modulation frequency and all odd-order harmonics to eliminate short-term noise resulting from RIN. However, we show that it is sufficient to suppress only certain frequency components of RIN, namely, around the modulation frequency and its third-order harmonic. This method is more effective for higher modulation frequency because it is not necessary to suppress RIN components over a wide range of frequencies. We demonstrate this method by direct feedback to the drive current of a superluminescent diode (SLD) having a center wavelength of 1.5\,$\mu$m.
Some portion of the SLD is picked up and detected by a photodetector, whose output voltage is used as a feedback signal to compensate for RIN.
Note that this method is not limited to an SLD but is applicable to other light sources. For example, using an acoustic optical modulator (AOM) after a light source, one can make feedback to the AOM to compensate for RIN similarly.

The short-term sensitivity is evaluated by the angular random walk~(ARW) coefficient.
Our simultaneous suppression of TPN and RIN yields an ARW of $\sim 15\,\mu\mathrm{deg}/\sqrt{\mathrm{h}}$, which is a 6-fold improvement compared to $\sim 93\,\mu\mathrm{deg}/\sqrt{\mathrm{h}}$ obtained by the conventional eigenfrequency modulation-demodulation technique.
Our FOG also shows good long-term stability, having a bias instability of $\sim 33\,\mu\mathrm{deg}/\mathrm{h}$ for a measurement time of 40 hours.

\section{Dominant noise factors and their suppression}

To improve the short-term sensitivity of a FOG, it is essential to examine the ratio of the FOG signal to the total noise of the system.
When a phase shift $\Delta \phi_\mathrm{R}$ is induced because of the Sagnac effect in the FOG, the detected signal photocurrent is described by
\begin{equation}
I = \frac{I_0}{2} \Big\{1+ \cos \left[ \Delta \phi_\mathrm{R} + \phi_\mathrm{m} \sin \omega_\mathrm{m} t \right] \Big\},
\label{eq:fog}
\end{equation}
where $\omega_\mathrm{m}$, $\phi_\mathrm{m}$, and $I_0$ represent the modulation frequency, modulation index, and signal current without rotation and modulation, respectively.
The corresponding rotation rate $\Omega_\mathrm{R}$ is calculated by $\Omega_\mathrm{R} = (c \lambda /4\pi RL) \Delta \phi_\mathrm{R} \equiv \Delta \phi_\mathrm{R}/S_\mathrm{F}$, where $c$, $\lambda$, $R$, $L$, and $S_\mathrm{F}$ are the speed of light, central wavelength,  radius of the fiber coil, fiber length, and scaling factor, respectively.
The output $I$ can be expanded using the harmonics of the modulation frequency, namely, $n\omega_\mathrm{m} \,(n=1,2,\cdots)$, and the corresponding Bessel functions $J_n (\phi_\mathrm{m})$ of integer order $n$ as follows:
\begin{eqnarray}
I & \approx & \frac{I_0}{2} \Big[1+ J_0 (\phi_\mathrm{m}) + 2 \sum_{k=1}^{\infty} J_{2k} (\phi_\mathrm{m}) \cos 2k\omega_\mathrm{m} t\Big] \nonumber \\
& &-I_0 \Delta \phi_\mathrm{R} \sum_{k=1}^{\infty} J_{2k-1} (\phi_\mathrm{m}) \sin (2k-1)\omega_\mathrm{m} t,
\label{eq:bess}
\end{eqnarray}
where we assume $\Delta \phi_\mathrm{R} \ll 1$.

The component of $I$ oscillating at the modulation frequency $\omega_\mathrm{m}$ is $I_\omega \approx I_0 \Delta \phi_\mathrm{R} J_1  (\phi_\mathrm{m})$.
We denote the demodulated current noises for TPN, RIN, shot noise, and detection noise by $\sigma_\mathrm{TPN}$, $\sigma_\mathrm{RIN}$, $\sigma_\mathrm{SN}$, and $\sigma_\mathrm{DN}$, respectively.
Eventually, an ARW is given by
\begin{equation}
\mathrm{ARW} = \frac{1}{S_\mathrm{F} J_1 (\phi_\mathrm{m}) I_0 \sqrt{B}} \sqrt{ \sigma^2_\mathrm{TPN} + \sigma^2_\mathrm{RIN} +\sigma^2_\mathrm{SN} +\sigma^2_\mathrm{DN}},
\label{eq:arw}
\end{equation}
where $B$ represents the detection bandwidth.

The TPN originates from the random thermal motion of silica particles in the optical fiber~\cite{Li2019,Knudsen1995,Moeller1996}.
This noise becomes more detrimental in a longer fiber.
The corresponding noise $\sigma_\mathrm{TPN}$ has been extensively studied~\cite{Li2019,Knudsen1995} and expressed by
\begin{eqnarray}
\sigma_\mathrm{TPN} & = & I_0 \bigg\{ \pi B \sum_{n=1}^{\infty} \Big[ J_{2n-1} (\phi_\mathrm{m}) - J_{2n+1} (\phi_\mathrm{m}) \Big]^2 \nonumber \\
& & \quad \times \langle \Delta \phi_{N,\mathrm{rms}}^2 (2n \omega_\mathrm{m}) \rangle \bigg\}^{1/2},
\label{eq:tpn}
\end{eqnarray}
where
\begin{eqnarray}
& & \langle \Delta \phi_{N,\mathrm{rms}}^2 (\omega) \rangle \nonumber \\
& & = \frac{k_B T^2 L}{\kappa \lambda^2} \left( \frac{\mathrm{d}n_\mathrm{eff}}{\mathrm{d}T}+n_\mathrm{eff} \alpha_L\right)^2 \times \ln \left[ \frac{\left(\frac{2}{W_0}\right)^4+\left(\frac{\omega}{D}\right)^2}{\left(\frac{4.81}{d}\right)^4+\left(\frac{\omega}{D}\right)^2}\right] \nonumber \\
& & \quad \times \left[ 1-\mathrm{sinc} \left(\frac{\omega L n_\mathrm{eff}}{c} \right) \right].
\end{eqnarray}
Here, $\langle \Delta \phi_{N,\mathrm{rms}}^2 (\omega) \rangle$ is the spectral density of phase noise introduced by the TPN.
$k_\mathrm{B}$ is Boltzmann's constant, and $\mathrm{d}n_\mathrm{eff}/\mathrm{d}T$ is the temperature coefficient of the effective refractive index $n_\mathrm{eff}$ of the fiber.
The fiber is characterized by mode field diameter $2W_0$, cladding diameter $d$, fiber length $L$, linear thermal expansion coefficient $\alpha_L$, thermal conductivity $\kappa$, and thermal diffusivity $D$.
Note that the expression in (\ref{eq:tpn}) differs from the equation in \cite{Li2019,Knudsen1995}, which considers that the spectral densities $\langle \Delta \phi_{N,\mathrm{rms}}^2 (\omega) \rangle$ and $\langle \Delta \phi_{N,\mathrm{rms}}^2 (-\omega) \rangle$ are statistically independent. However, they are correlated, and our equation accounts for this fact~\cite{MM2022}.
Based on this expression, the effect of TPN can be reduced by increasing the modulation frequency, which corresponds to the odd-order harmonic of the eigenfrequency of the coil.
We apply this approach to our coil made of a polarization-maintaining~(PM) fiber with $L \sim 4920$\,m below.

Excess RIN is another dominant noise originating from the random beating between all the frequency components within the broad spectrum of a light source~\cite{Burns1990,Burns1996,Shin2010,Blake1994}.
The amount of noise is simply given by the inverse of the frequency spectrum linewidth $\Delta \nu$ of the light source. 
In the lock-in measurement, the RIN is distributed between the signal and quadrature phases, and the signal phase component $\sigma_\mathrm{RIN}$ can be calculated as follows:
\begin{eqnarray}
\sigma_\mathrm{RIN}& = &\frac{I_0 \sqrt{B}}{2 \sqrt{\Delta \nu} } \bigg\{ \Bigl[ 1+J_0 (\phi_\mathrm{m})-J_2 (\phi_\mathrm{m}) \Bigr]^2 \nonumber \\
& & +\Bigl[J_{2} (\phi_\mathrm{m})-J_{4} (\phi_\mathrm{m}) \Bigr]^2 \nonumber \\
& & + \Bigl[J_{4} (\phi_\mathrm{m})-J_{6} (\phi_\mathrm{m}) \Bigr]^2 + \cdots \bigg\}^{1/2},
\label{eq:rin}
\end{eqnarray}
under the assumption that the RIN is white and each of its spectral components is statistically independent~\cite{Blake1994}.
The spectrum bandwidth $\Delta \nu$ of the light source can be calculated using the expression
\begin{equation}
\Delta \nu =\frac{\left[ \int P(\nu) d\nu\right]^2}{\int P^2(\nu) d\nu}, 
\label{eq:dnu}
\end{equation}
where $P(\nu)$ is the power spectral density of the light source with respect to frequency $\nu$~\cite{Burns1990}.
When RIN is the dominant noise source, the ARW becomes a minimum at $\phi_\mathrm{m} \approx 2.7$~\cite{Blake1994}.
Note that the $n$-th term in the square root in (\ref{eq:rin}) is originated by the mixing between the RIN at frequency $(2n-1)\,\omega_\mathrm{m}$ and the modulated signal at frequencies $[(2n-1)\pm 1]\,\omega_\mathrm{m}$. Thus, complete elimination of $\sigma_\mathrm{RIN}$ requires the suppression of all odd-order $\omega_\mathrm{m}$ components of RIN. However, as the first and second terms in (\ref{eq:rin}) are bigger than the subsequent terms, the suppression of only $\omega_\mathrm{m}$ and $3\,\omega_\mathrm{m}$ components of RIN results in a significant reduction of $\sigma_\mathrm{RIN}$.

When the dominant noises such as TPN and RIN are sufficiently suppressed, shot noise, which is given by
\begin{eqnarray}
\sigma_\mathrm{SN} & = & \bigg\{e B I_0 \sum_{s=0}^{\infty} \left[ J_{2s}\left(\frac{\phi_\mathrm{m}}{2}\right) - J_{2(s+1)}\left(\frac{\phi_\mathrm{m}}{2}  \right) \right]^2 \bigg\}^{1/2}   \nonumber \\
& = & \bigg\{\frac{e B I_0}{2} \left[ 1+J_0 (\phi_\mathrm{m})-J_2 (\phi_\mathrm{m}) \right] \bigg\}^{1/2},
\label{eq:pnsn}
\end{eqnarray}
and detection noise will finally limit the sensitivity.
Note that the expression in (\ref{eq:pnsn}) differs from the conventional equation, which only includes the effects of the DC component in the detected signal~\cite{Burns1990,Moeller1989}. In contrast, our equation also accounts for the AC contribution, which modifies the amount of noise in the case of a lock-in measurement~\cite{MM2022}.

Detection noise is generated mainly from a transimpedance amplifier with a photodiode~\cite{Scott2001} and a lock-in amplifier, which is given by
\begin{equation}
\sigma_\mathrm{DN} = \sqrt{\frac{B}{2}} \bigg\{2 e i_\mathrm{D} + i_\mathrm{NT}^2 + \left( \frac{v_\mathrm{NT}}{R_\mathrm{F}}\right)^2 + \frac{4k_\mathrm{B}T}{R_\mathrm{F}}+\left( \frac{v_\mathrm{NL}}{R_\mathrm{F}} \right)^2 \bigg\}^{1/2},
\label{eq:pndn}
\end{equation}
where $i_\mathrm{D}$ is the dark current of the photodiode, $i_\mathrm{NT}$ and $v_\mathrm{NT}$ are the input current noise spectral density (NSD) and input voltage NSD of a transimpedance amplifier, $R_\mathrm{F}$ is the feedback resistance, and $v_\mathrm{NL}$ is the input voltage NSD of a lock-in amplifier.

\begin{figure}[t]
\centering\includegraphics[width=8.8cm]{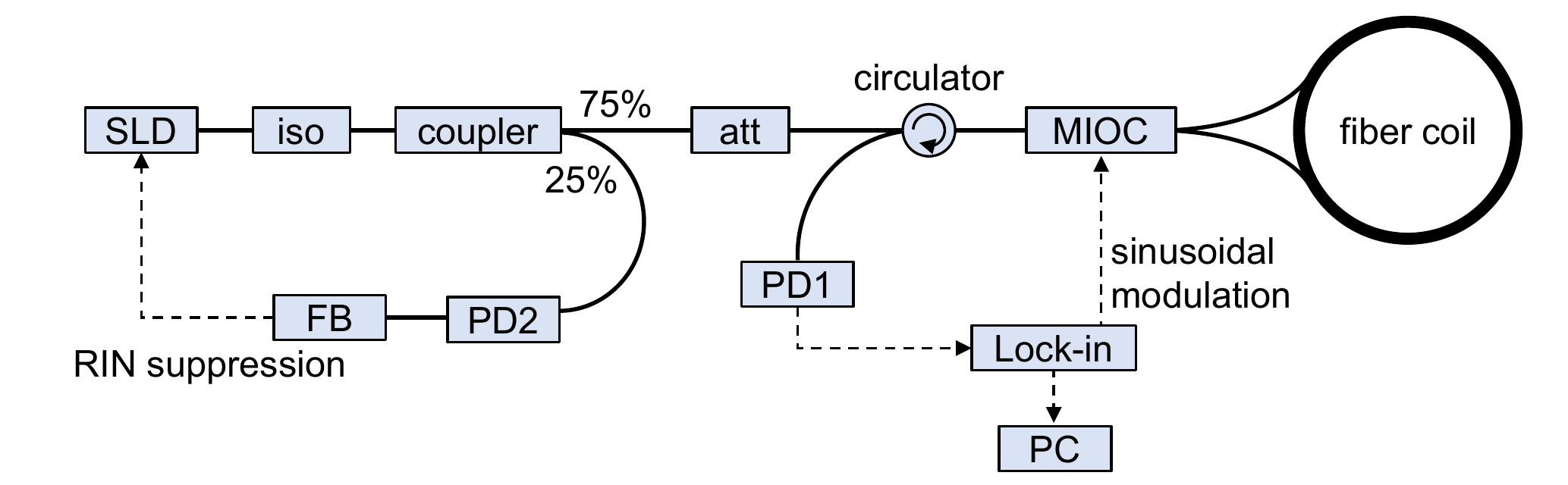}
\caption{Schematic setup of the FOG apparatus. SLD, superluminescent diode; iso, isolator; att, variable attenuator; MIOC, multifunctional integrated optical chip; PD, photodiode; PC, personal computer; FB, feedback circuit.} 
\label{fig:setup}
\end{figure}

\begin{figure}[t]
\centering\includegraphics[width=7.5cm]{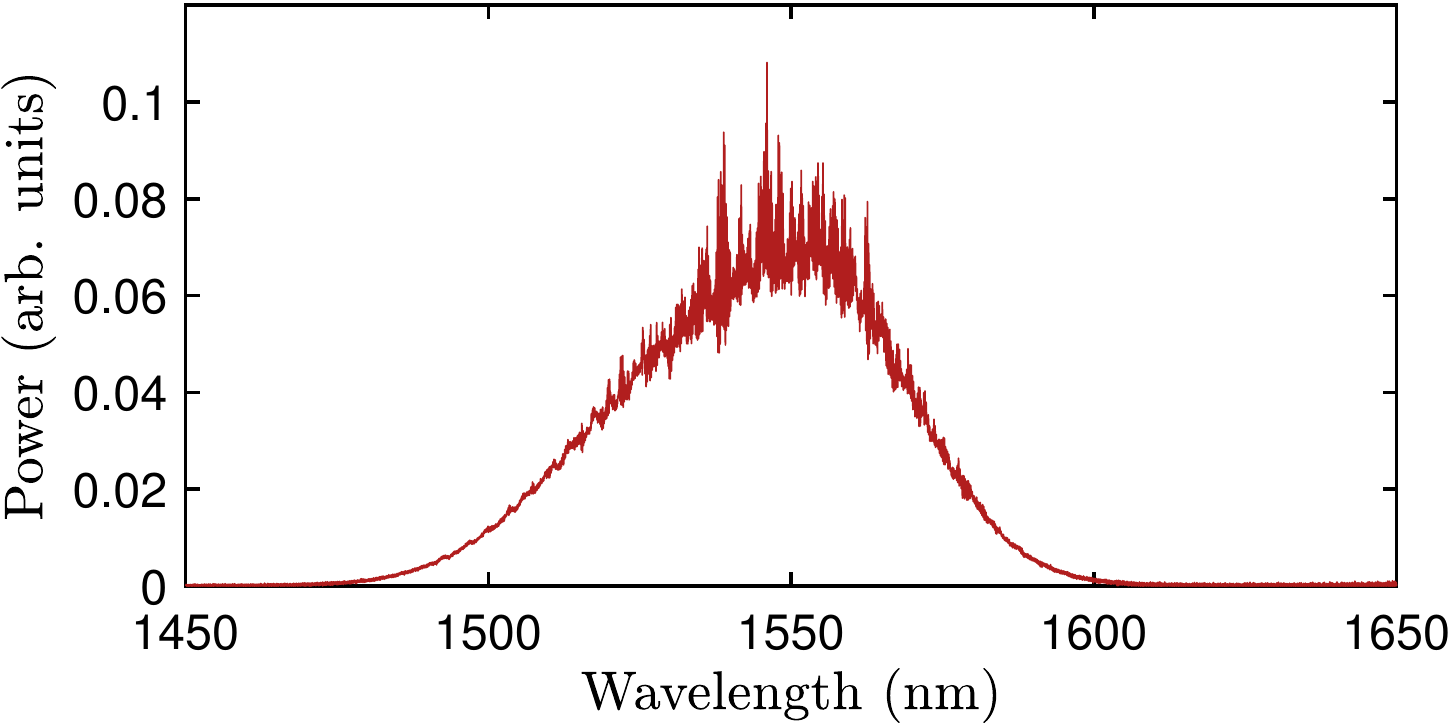}
\caption{Measured spectrum of the SLD at the photodetector PD1.} 
\label{fig:spectrum}
\end{figure}

\section{Experimental setup}
Our all-fiber-based experimental apparatus is schematically shown in Fig.~\ref{fig:setup}. The light source is an SLD with a central wavelength $\lambda$ of $\sim 1544$\,nm and a bandwidth of 52\,nm (full width at half maximum) (see Fig.~\ref{fig:spectrum}).
The central wavelength is calculated as an expectation value of the spectrum distribution in Fig.~\ref{fig:spectrum}.
The SLD output goes through an isolator to avoid back reflection.
Next, the primary output from the PM fiber coupler goes to a standard setup of a FOG, which consists of a circulator, a multifunctional integrated optical chip (MIOC), and a fiber coil. 
The fiber coil is produced by winding a PM fiber with a length $L$ of $\sim 4920$\,m and an average radius $R$ of 115\,mm using a quadruple winding pattern.
Its eigenfrequency $f_\mathrm{e}$ is estimated to be $20.3$\,kHz.
Sinusoidal modulation is applied to the MIOC.
The output from the circulator is detected by an InGaAs photodiode (PD1) with a sensitivity of 0.95\,A/W, 
followed by a transimpedance amplifier with a gain of 12\,kV/A. The optical attenuation in the coil, including the MIOC, is measured to be 16.3\,dB. 
This includes losses due to inherent fiber attenuation ($\sim$6.9\,dB) and fabrication process, such as winding-induced fiber bending loss and stress-induced loss.
The typical power without phase modulation is 450\,$\mu$W at PD1.
The detected signal is measured by a lock-in amplifier.
The minor output from the coupler is used for RIN suppression. The signal detected by a photodiode (PD2) is sent to a homemade feedback circuit, whose output voltage is fed back to the current controller of the SLD. The details are mentioned later.

As shown in (\ref{eq:bess}), in the case of sinusoidal phase modulation, the output current $I$ can be expressed using the harmonics of the modulation frequency and the Bessel functions.
The magnitudes of the $i$-th harmonics $h_i$ can be measured by a lock-in amplifier with a multifrequency demodulation function.
Using the ratio of the first and second harmonic components $h_1$ and $h_2$, the Sagnac phase shift $\Delta \phi_\mathrm{R}$ can be obtained in the following form~\cite{Bohm1983}:
\begin{equation}
\Delta \phi_\mathrm{R} =\tan^{-1} \left( \frac{J_2 (\phi_\mathrm{m} )}{J_1 (\phi_\mathrm{m} )} \frac{h_1}{h_2} \right).
\end{equation}
Here the information of $\phi_\mathrm{m}$ is extracted from the following relation:
\begin{equation}
\frac{h_2}{h_4} = \frac{J_2 (\phi_\mathrm{m} )}{J_4 (\phi_\mathrm{m} )},
\end{equation}
where $h_4$ is the magnitude of the fourth-order harmonic component.
Note that the fluctuation of the SLD power and that of the modulation index can be compensated for by this method.

The fiber coil and MIOC are placed in a vacuum chamber with pressure below $5\times10^{-2}$\,Pa, which is surrounded by a magnetic shield and temperature stabilized at 24.0 $^\circ \mathrm{C}$ within 1\,mK.

\section{Results}

\begin{figure}[t]
\centering\includegraphics[width=8.8cm]{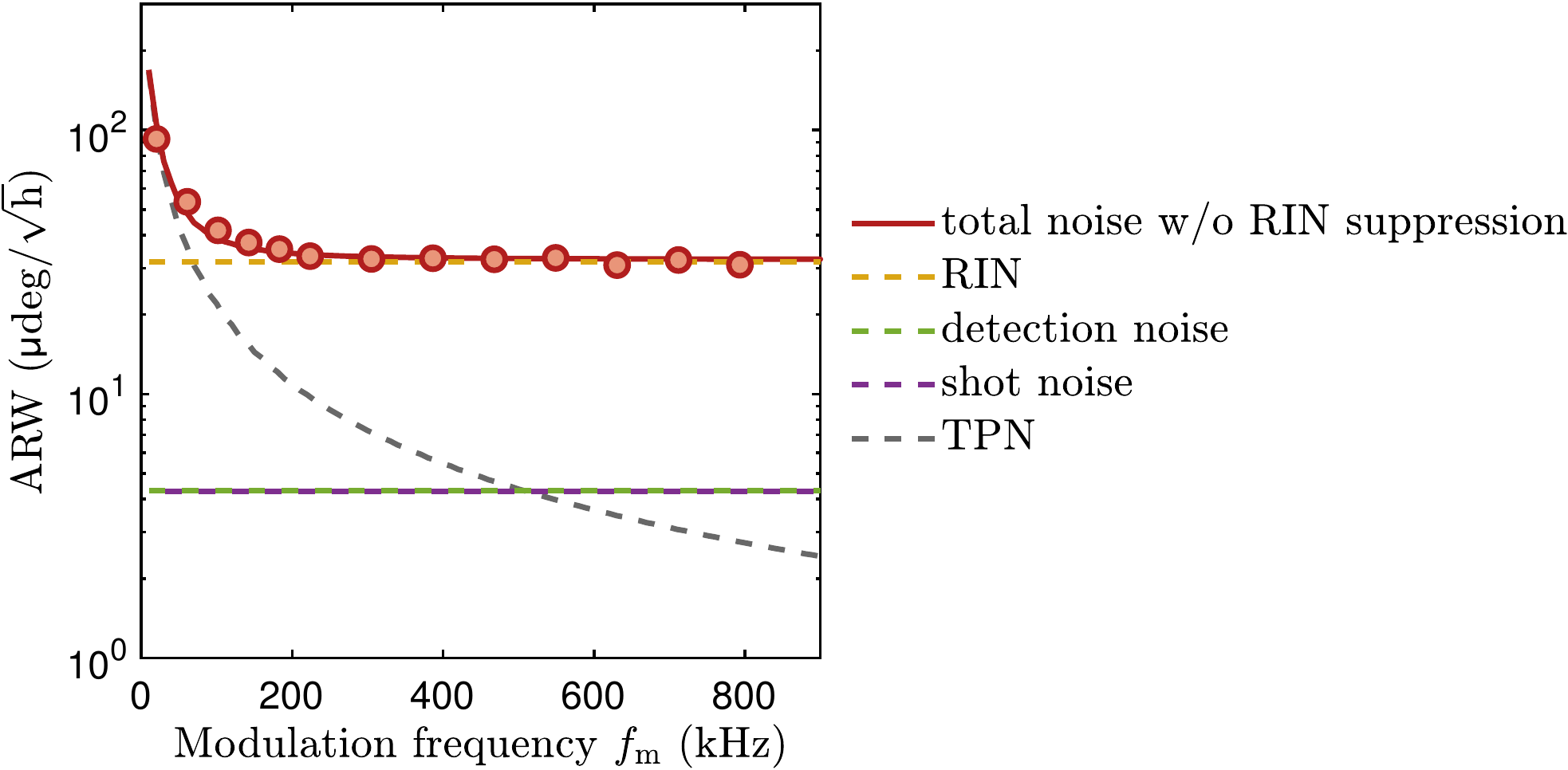}
\caption{Modulation frequency dependence of an angular random walk (ARW). Measured values of an ARW are plotted as a function of modulation frequency $f_\mathrm{m} = \omega_\mathrm{m}/2\pi$ for the case without RIN suppression (red circle points).
The red solid line shows the fitted result of the total noise. Calculation results for RIN and TPN using the fitted values are presented by orange and grey dashed lines, respectively.
The purple and green dashed lines show the calculations for shot noise and detection noise, respectively.
The modulation depth is set to $\phi_\mathrm{m}= 2.7$.
The error bars of the data points, which are the standard deviation of the fitting of the ARW, are smaller than the marker size.
} 
\label{fig:arw}
\end{figure}

Fig.~\ref{fig:arw} shows the measured ARW (red circle data points) as a function of modulation frequency $f_\mathrm{m}=\omega_\mathrm{m}/2\pi$.
Here, we measure the Allan deviation of the angular rate $\Omega_\mathrm{R}$ as a function of integration time $\tau$ and fit it with a function $\mathrm{ARW}/\sqrt{\tau}$. 
Note that excess RIN is not yet suppressed at this stage, and we use the modulation index $\phi_\mathrm{m} = 2.7$ where the optimal ARW can be expected.
The measured values of ARW decrease monotonically with the modulation frequency and reach a lower limit of $\sim 31\,\mathrm{\mu deg}/\sqrt{\mathrm{h}}$ in a frequency range larger than $386$\, kHz, which corresponds to the 19-th order harmonic of the eigenfrequency $f_\mathrm{e}$ in our setting.

\begin{figure}[t]
\centering\includegraphics[width=8.8cm]{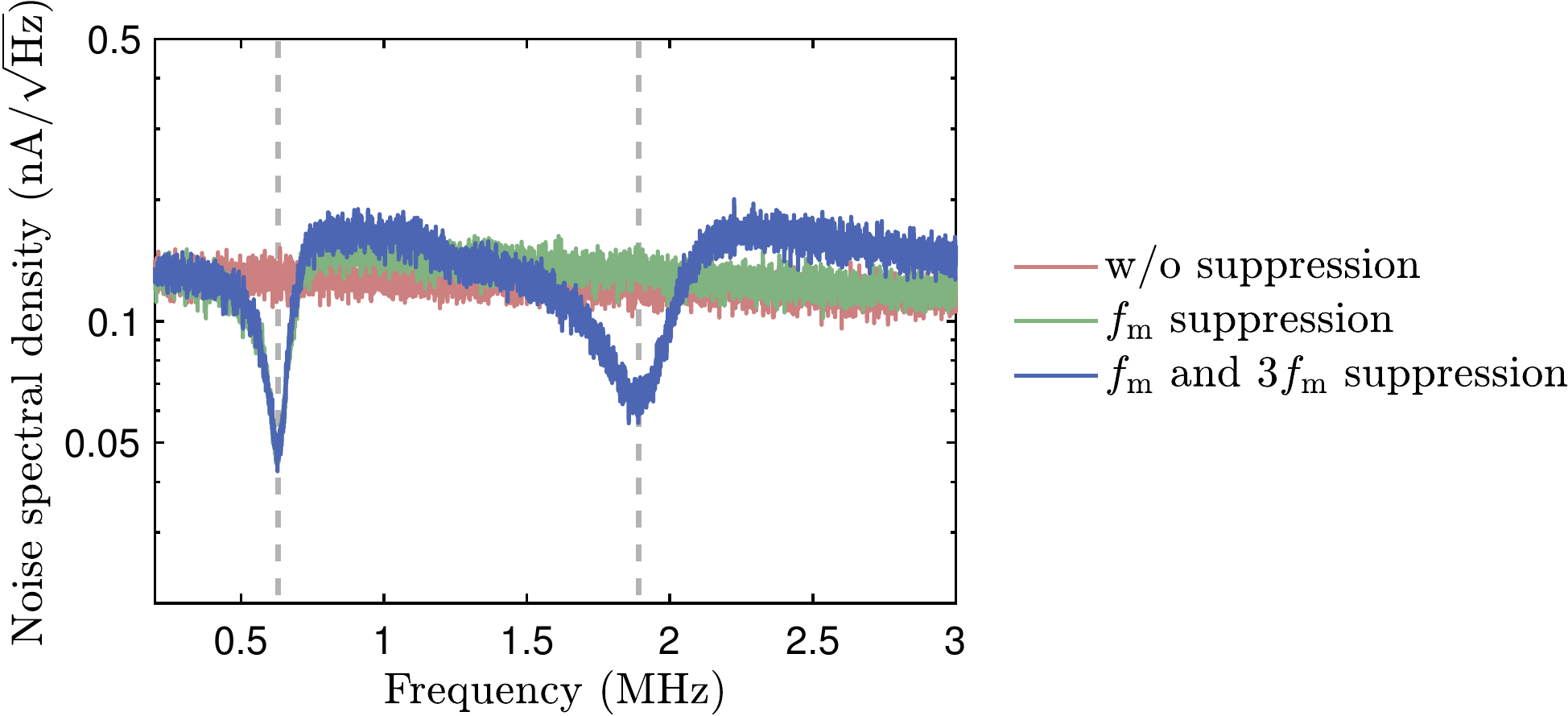}
\caption{
Noise spectral densities of the light source measured at photodiode PD1.
The red trace shows the result without RIN suppression.
The green and blue traces show the results with RIN suppression at $f_\mathrm{m}$ only and at $f_\mathrm{m}$ and $3f_\mathrm{m}$, respectively.
The vertical dashed lines indicate $f_\mathrm{m}$ and $3f_\mathrm{m}$.
} 
\label{fig:nd}
\end{figure}

This figure also shows a fitting curve to the measured data points with (\ref{eq:arw}), which considers TPN, RIN, shot noise, and detection noise.
To calculate $\sigma_\mathrm{TPN}$, we use the values written in \cite{Li2019} regarding the fiber characteristics such as $n_\mathrm{eff}$, $\mathrm{d}n_\mathrm{eff}/\mathrm{d}T$, $\alpha_L$, and $\kappa$.
Other parameters are $2W_0= 6.7\,\mu$m and $d=80\,\mu$m from our manufacturer's data sheet.
We use the thermal diffusivity of an optical fiber as a fitting parameter and obtain a value of $(5.5\pm0.8)\times 10^{-7} \mathrm{m}^2 \mathrm{s}^{-1}$, which has no discrepancy with the values in the literature~\cite{Li2019,Knudsen1995,Moeller1996,Chardon1983}. 
The excess RIN is calculated by (\ref{eq:rin}), and the spectrum bandwidth $\Delta \nu$ of the light source in (\ref{eq:dnu}) is obtained as $\Delta \nu = 9.71$\,THz from the measured spectrum in Fig.~\ref{fig:spectrum}.
This model assumes pure spontaneous emission as a light source, but due to the amplification process in the SLD, the experimental value of the RIN can be smaller than the calculated one~\cite{Suo2021}.
Therefore, the previous work introduced the noise suppression factor $\eta$ and divided the noise $\sigma_\mathrm{RIN}$ by $\sqrt{\eta}$~\cite{Shin2010}.
Similarly, we use this factor $\eta$ as a fitting parameter and obtain $\eta = 1.91\pm 0.17$, which is comparable to the values in \cite{Shin2010}. 
The magnitudes of shot noise in (\ref{eq:pnsn}) and detection noise in (\ref{eq:pndn}) are calculated with our experimental parameters.
The contributions of RIN, TPN, shot noise, and detection noise are plotted in Fig.~\ref{fig:arw} as dashed lines.

Clearly, in terms of the TPN suppression, a higher modulation frequency is better: in the lower modulation-frequency range below $100$\,kHz, the TPN is dominant, whereas its contribution decreases in the higher frequency range, and the RIN becomes dominant.
Therefore, suppressing the RIN in the high frequency range can further improve the short-term sensitivity of the FOG. On the other hand, there are technical difficulties in reducing RIN at a higher frequency. We need a photo-detector and lock-in amplifier with large bandwidth and also a high-speed feedback circuit. In view of these conditions, hereafter we set our modulation frequency $f_\mathrm{m}$ to $630.6$\,kHz ($f_\mathrm{m}=31f_\mathrm{e}$), at which the TPN contributes just less than the shot noise.

As discussed in the theoretical section, we suppress excess RIN at particular frequencies, namely, $f_\mathrm{m}$ and/or $3f_\mathrm{m}$, using the feedback to the current driver of the SLD.
As shown in Fig.~1, some portion of the SLD output is picked off by the fiber coupler and detected by PD2.
The output voltage of the transimpedance amplifier after PD2 is split into two.
Each of the outputs goes through a bandpass filter whose resonance frequency is set to $f_\mathrm{m}$ or $3f_\mathrm{m}$, and its output is used in the feedback.
Fig.~\ref{fig:nd} shows the noise spectral density detected by PD1 when phase modulation to the MIOC is not applied.
The RIN is successfully suppressed around the frequencies $f_\mathrm{m}$ and $3f_\mathrm{m}$ by a factor of $\sim$2.7 and $\sim$2.1, respectively.

\begin{figure}[!t]
\centering\includegraphics[width=8.8cm]{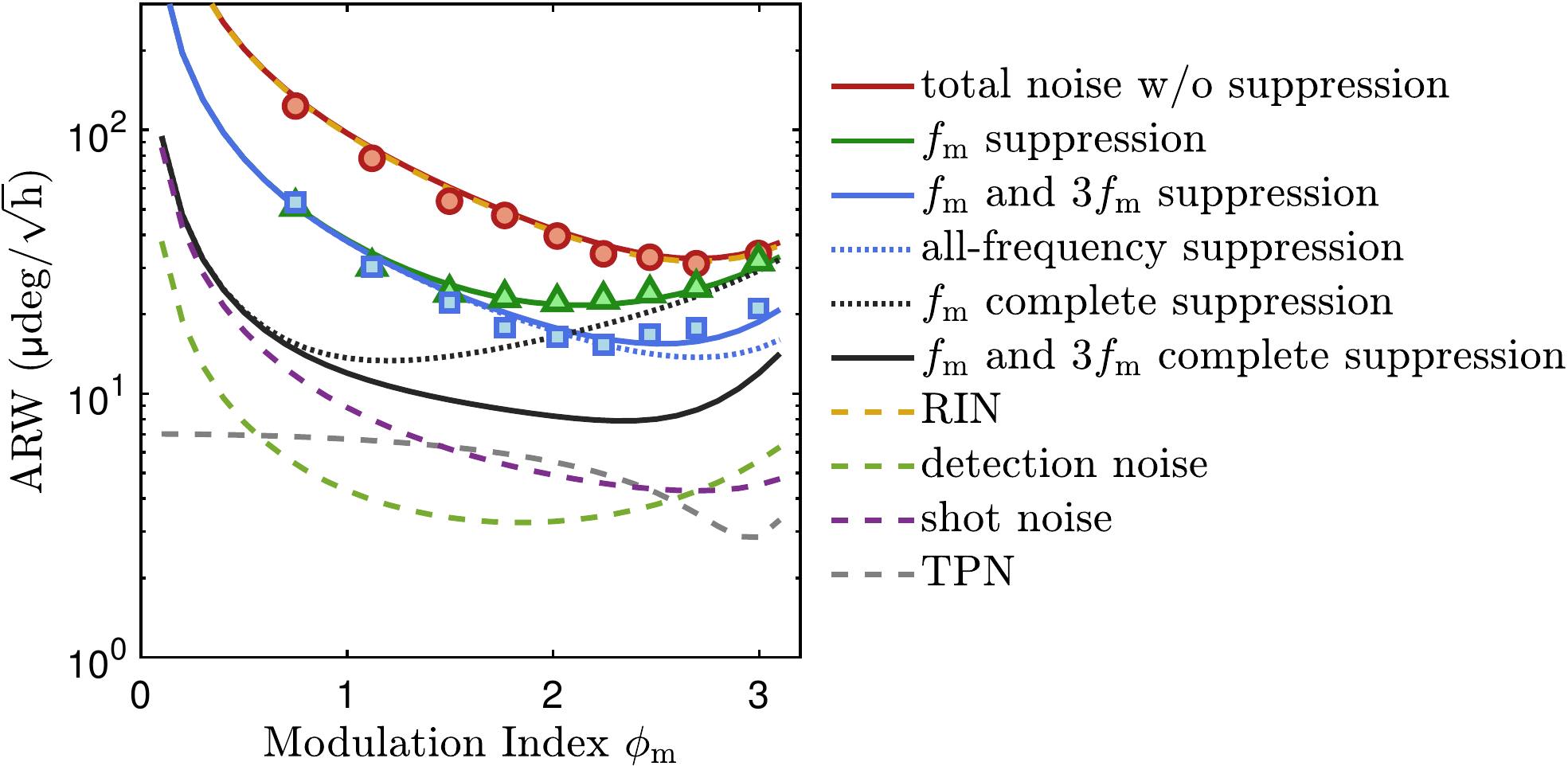}
\caption{Modulation index dependence of an ARW at $f_\mathrm{m}=630.6$\,kHz.
Measured values of an ARW are plotted for the cases without RIN suppression (red circle points) and with RIN suppression at $f_\mathrm{m}$ only (green triangle points) and at $f_\mathrm{m}$ and $3f_\mathrm{m}$ (blue square points).
The red, green, and blue solid lines are the corresponding theoretical calculations using the fitted values obtained in Fig.~\ref{fig:arw}.
The blue dotted line is the theoretical calculation assuming that all the frequency components of the RIN are suppressed by the same amount as the $f_\mathrm{m}$ component.
The black dotted and solid lines correspond to ARW values when the RIN is completely suppressed at $f_\mathrm{m}$ only and at $f_\mathrm{m}$ and $3f_\mathrm{m}$, respectively.
The orange, green, purple, and grey dashed lines show the calculated results for the RIN, detection noise, shot noise, and TPN, respectively.
The error bars of the data points, which are the standard deviation of the fitting of the ARW, are smaller than the marker size.} 
\label{fig:arw31}
\end{figure}

Under these RIN-suppression conditions, we measure the dependence of the ARW on the modulation index $\phi_\mathrm{m}$, as shown in Fig.~\ref{fig:arw31}. 
First, we examine the case without RIN suppression.
The red circle points in Fig.~\ref{fig:arw31} are the measured results, showing a minimum value of $\sim 31\,\mu\mathrm{deg}/\sqrt{\mathrm{h}}$ at $\phi_\mathrm{m} \approx 2.7$, as expected.
We also measure the case with RIN suppression at $f_\mathrm{m}$ only.
The green triangle points are the measured results, clearly showing that the ARW values are significantly improved compared to the case without RIN suppression.
The minimum value in this case is $\sim 22\,\mu\mathrm{deg}/\sqrt{\mathrm{h}}$, and its position is shifted to $\phi_\mathrm{m} \approx 2.0$.
Similarly, the measured results for the case of RIN suppression at $f_\mathrm{m}$ and $3f_\mathrm{m}$ are presented by the blue square points. 
We obtain the minimum ARW value $\sim 15\,\mu\mathrm{deg}/\sqrt{\mathrm{h}}$ at $\phi_\mathrm{m} = 2.24$. All the sets of measured results are in good agreement with their corresponding theoretical curves, which suggests that our understanding of the RIN based on (\ref{eq:rin}) is correct.
This understanding shows that even if all the frequency components of the RIN are suppressed by the same amount as the $f_\mathrm{m}$ component ($\sim$2.7), the ARW value does not improve significantly (the blue dotted line).
Achievable ARW remains at the same level ($\sim 13\,\mu\mathrm{deg}/\sqrt{\mathrm{h}}$), even if we suppress only the $f_\mathrm{m}$ component entirely (the black dotted line). However, drastic improvement can be obtained by completely suppressing $f_\mathrm{m}$ and $3f_\mathrm{m}$ components, reaching a value of $\sim 8\,\mu\mathrm{deg}/\sqrt{\mathrm{h}}$ (the black solid line).

\begin{figure}[!t]
\centering\includegraphics[width=7.5cm]{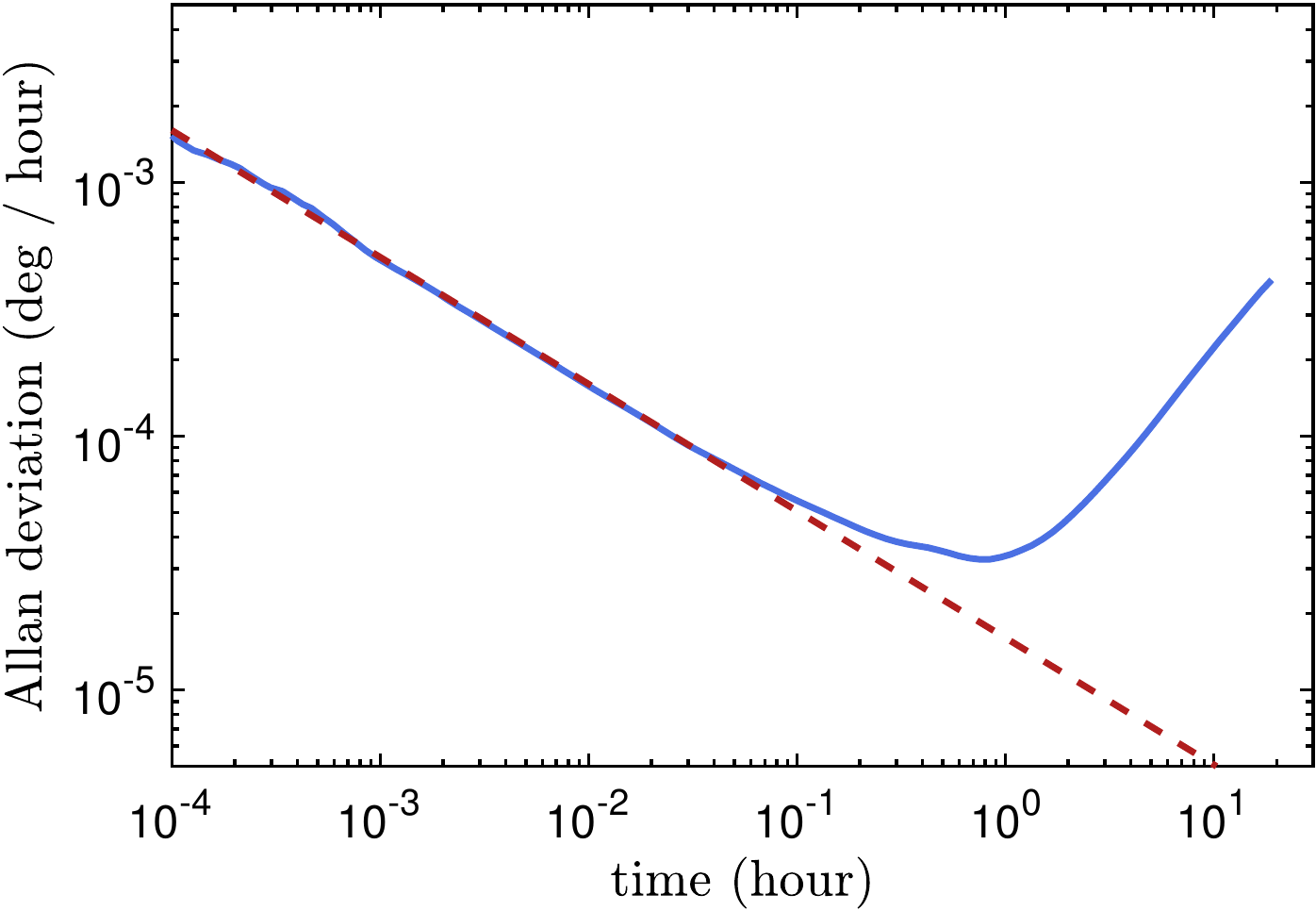}
\caption{Allan deviation of the measured rotation rate $\Omega_\mathrm{R}$. The experimental condition is the same as that corresponding to the blue square point at $\phi_\mathrm{m} \approx 2.24$ in Fig.~\ref{fig:arw31}.
The red dashed line shows the fitted result with a function of $\mathrm{ARW}/\sqrt{\tau}$.
}
\label{fig:allan}
\end{figure}

We also examine the long-term stability using the optimum condition mentioned above.
Fig.~\ref{fig:allan} shows the measured Allan deviation of the rotation rate $\Omega_\mathrm{R}$ for a measurement time of 40 hours.
Despite such a long measurement time, we achieve a bias stability of $\sim 33\,\mu\mathrm{deg/h}$, showing that our FOG has excellent long-term stability.
For a time scale longer than several hours, a ramp rate with a slope of $+1$ is observed, indicating a very slow monotonic change in the FOG signal over a long period of time~\cite{IEEE1998}.
This change could be due to a drift of the central wavelength and the corresponding scale factor.
An investigation of the ramp rate is necessary but beyond the scope of the present study.
Please note that we use a temperature-controlled environment in the vacuum chamber with a magnetic shield to aim to obtain long-term stability, whereas the essential point in our method of simultaneous suppression of TPN and RIN is the improvement of short-term sensitivity. We verified that this method works also at room temperature and in the atmosphere.

The ARW of $15\,\mu\mathrm{deg}/\sqrt{\mathrm{h}}$ corresponds to an interferometer phase noise of $7\times 10^{-8}\,\mathrm{rad}/\sqrt{\mathrm{Hz}}$ and provides a root mean square phase noise of $7\times 10^{-9}\,\mathrm{rad}$ after $100$ seconds of average time. The total phase accumulated after the propagation in the fiber coil is $3\times 10^{10}\,\mathrm{rad}$. Therefore, reciprocity is estimated to be $\sim 2\times10^{-19}$, which is the same order as the current high-performance FOG~\cite{Lefevre2020}.

\begin{table}[!ht]
\caption{Parameters of high-performance interferometric FOGs}
\label{table:ARW}
\centering
\begin{tabular}{c c c c c}
\hline
ARW & fiber length & effective area & Ref.\\
\hline
$69\,\mu\mathrm{deg}/\sqrt{\mathrm{h}}$ & $5000$\,m & $310\,\mathrm{m}^2$ &  \cite{Korkishko2017}\\
$32\,\mu\mathrm{deg}/\sqrt{\mathrm{h}}$ & $5000$\,m & $230\,\mathrm{m}^2$ & \cite{Lefevre2020}\\
$12\,\mu\mathrm{deg}/\sqrt{\mathrm{h}}$ & $30630$\,m & $2900\,\mathrm{m}^2$ & \cite{Li2019}\\
$15\,\mu\mathrm{deg}/\sqrt{\mathrm{h}}$ & $4920$\,m & $280\,\mathrm{m}^2$ & present work\\
\hline
\end{tabular}
\end{table}

\section{Conclusion}

We have demonstrated the simultaneous suppression of TPN and RIN. 
We have revisited the derivation of the RIN distribution in the lock-in detection scheme and shown that it is sufficient to suppress the RIN components at only the modulation frequency and its third-order harmonic. 
We observed significant improvement of the ARW coefficient by simply suppressing the first- and third-order components, where we employed the modulation frequency of the 31st harmonics of the eigenfrequency of the fiber coil.
Although suppression of RIN at all frequency components is technically difficult, especially for coils with large eigenfrequencies, suppression of only two low-frequency components is relatively simple.
Our simultaneous suppression of TPN and RIN yielded an ARW coefficient of $\sim 15\,\mu\mathrm{deg}/\sqrt{\mathrm{h}}$, as well as a bias instability of $\sim 33\,\mu\mathrm{deg}/\mathrm{h}$ for a measurement time of 40 hours.
This ARW is, to our knowledge, the best one in the world for FOGs using a five-kilometers long fiber coil as represented in Table~\ref{table:ARW}.
Achieving such a small ARW is very important in practical INS applications, reducing the time required for initial INS alignment.
As discussed in~\cite{Paturel2014}, bias instability below $20\,\mu\mathrm{deg}/\mathrm{h}$ is necessary to maintain a navigational accuracy of a one nautical mile per month.
In order to reach this level of accuracy within one-hour alignment, the ARW must be less than $20\,\mu\mathrm{deg}/\sqrt{\mathrm{h}}$~\cite{Paturel2014}.
Our performance has reached this level, and therefore paves the way for practical application to ultra-precise inertial navigation in long-term operation.

\section*{Acknowledgments}
We thank K. Aikawa, Y. Kamino, H. Inaba, Y. Yatsu, T. Mukaiyama, D. Akamatsu, T. Ido, M. Yasuda, and K. Tsuchiya for fruitful discussions. We are grateful to H. Ishijima for his technical assistance with the experiments.
This work was supported by JST, Grant Number JPMJPF2015 and JPMJMI17A3.


\bibliographystyle{apsrev4-2}
\bibliography{simul_supp.bib}

\begin{thebibliography}{34}%
\makeatletter
\providecommand \@ifxundefined [1]{%
 \@ifx{#1\undefined}
}%
\providecommand \@ifnum [1]{%
 \ifnum #1\expandafter \@firstoftwo
 \else \expandafter \@secondoftwo
 \fi
}%
\providecommand \@ifx [1]{%
 \ifx #1\expandafter \@firstoftwo
 \else \expandafter \@secondoftwo
 \fi
}%
\providecommand \natexlab [1]{#1}%
\providecommand \enquote  [1]{``#1''}%
\providecommand \bibnamefont  [1]{#1}%
\providecommand \bibfnamefont [1]{#1}%
\providecommand \citenamefont [1]{#1}%
\providecommand \href@noop [0]{\@secondoftwo}%
\providecommand \href [0]{\begingroup \@sanitize@url \@href}%
\providecommand \@href[1]{\@@startlink{#1}\@@href}%
\providecommand \@@href[1]{\endgroup#1\@@endlink}%
\providecommand \@sanitize@url [0]{\catcode `\\12\catcode `\$12\catcode
  `\&12\catcode `\#12\catcode `\^12\catcode `\_12\catcode `\%12\relax}%
\providecommand \@@startlink[1]{}%
\providecommand \@@endlink[0]{}%
\providecommand \url  [0]{\begingroup\@sanitize@url \@url }%
\providecommand \@url [1]{\endgroup\@href {#1}{\urlprefix }}%
\providecommand \urlprefix  [0]{URL }%
\providecommand \Eprint [0]{\href }%
\providecommand \doibase [0]{https://doi.org/}%
\providecommand \selectlanguage [0]{\@gobble}%
\providecommand \bibinfo  [0]{\@secondoftwo}%
\providecommand \bibfield  [0]{\@secondoftwo}%
\providecommand \translation [1]{[#1]}%
\providecommand \BibitemOpen [0]{}%
\providecommand \bibitemStop [0]{}%
\providecommand \bibitemNoStop [0]{.\EOS\space}%
\providecommand \EOS [0]{\spacefactor3000\relax}%
\providecommand \BibitemShut  [1]{\csname bibitem#1\endcsname}%
\let\auto@bib@innerbib\@empty
\bibitem [{\citenamefont {Lef{\`e}vre}(2022)}]{Lefevre2014}%
  \BibitemOpen
  \bibfield  {author} {\bibinfo {author} {\bibfnamefont {H.~C.}\ \bibnamefont
  {Lef{\`e}vre}},\ }\href@noop {} {\emph {\bibinfo {title} {The fiber-optic
  gyroscope}}},\ \bibinfo {edition} {3rd}\ ed.\ (\bibinfo  {publisher} {Artech
  house},\ \bibinfo {year} {2022})\BibitemShut {NoStop}%
\bibitem [{\citenamefont {Lef{\`e}vre}\ \emph {et~al.}(2020)\citenamefont
  {Lef{\`e}vre} \emph {et~al.}}]{Lefevre2020}%
  \BibitemOpen
  \bibfield  {author} {\bibinfo {author} {\bibfnamefont {H.~C.}\ \bibnamefont
  {Lef{\`e}vre}} \emph {et~al.},\ }in\ \href@noop {} {\emph {\bibinfo
  {booktitle} {Optical Waveguide and Laser Sensors}}},\ Vol.\ \bibinfo {volume}
  {11405}\ (\bibinfo  {publisher} {SPIE},\ \bibinfo {year} {2020})\ pp.\
  \bibinfo {pages} {10 -- 29}\BibitemShut {NoStop}%
\bibitem [{\citenamefont {Korkishko}\ \emph {et~al.}(2017)\citenamefont
  {Korkishko}, \citenamefont {Fedorov}, \citenamefont {Prilutskiy},
  \citenamefont {Ponomarev}, \citenamefont {Fedorov}, \citenamefont
  {Kostritskii}, \citenamefont {Morev}, \citenamefont {Obuhovich},
  \citenamefont {Prilutskiy}, \citenamefont {Zuev},\ and\ \citenamefont
  {Varnakov}}]{Korkishko2017}%
  \BibitemOpen
  \bibfield  {author} {\bibinfo {author} {\bibfnamefont {Y.~N.}\ \bibnamefont
  {Korkishko}}, \bibinfo {author} {\bibfnamefont {V.~A.}\ \bibnamefont
  {Fedorov}}, \bibinfo {author} {\bibfnamefont {V.~E.}\ \bibnamefont
  {Prilutskiy}}, \bibinfo {author} {\bibfnamefont {V.~G.}\ \bibnamefont
  {Ponomarev}}, \bibinfo {author} {\bibfnamefont {I.~V.}\ \bibnamefont
  {Fedorov}}, \bibinfo {author} {\bibfnamefont {S.~M.}\ \bibnamefont
  {Kostritskii}}, \bibinfo {author} {\bibfnamefont {I.~V.}\ \bibnamefont
  {Morev}}, \bibinfo {author} {\bibfnamefont {D.~V.}\ \bibnamefont
  {Obuhovich}}, \bibinfo {author} {\bibfnamefont {S.~V.}\ \bibnamefont
  {Prilutskiy}}, \bibinfo {author} {\bibfnamefont {A.~I.}\ \bibnamefont
  {Zuev}},\ and\ \bibinfo {author} {\bibfnamefont {V.~K.}\ \bibnamefont
  {Varnakov}},\ }in\ \href@noop {} {\emph {\bibinfo {booktitle} {2017 DGON
  Inertial Sensors and Systems (ISS)}}}\ (\bibinfo {year} {2017})\ pp.\
  \bibinfo {pages} {1--23}\BibitemShut {NoStop}%
\bibitem [{\citenamefont {Lef{\`e}vre}(2016)}]{Lefevre2016}%
  \BibitemOpen
  \bibfield  {author} {\bibinfo {author} {\bibfnamefont {H.~C.}\ \bibnamefont
  {Lef{\`e}vre}},\ }in\ \href@noop {} {\emph {\bibinfo {booktitle} {Fiber Optic
  Sensors and Applications XIII}}},\ Vol.\ \bibinfo {volume} {9852},\ \bibinfo
  {editor} {edited by\ \bibinfo {editor} {\bibfnamefont {E.}~\bibnamefont
  {Udd}}, \bibinfo {editor} {\bibfnamefont {G.}~\bibnamefont {Pickrell}},\ and\
  \bibinfo {editor} {\bibfnamefont {H.~H.}\ \bibnamefont {Du}}},\ \bibinfo
  {organization} {International Society for Optics and Photonics}\ (\bibinfo
  {publisher} {SPIE},\ \bibinfo {year} {2016})\ pp.\ \bibinfo {pages} {9 --
  18}\BibitemShut {NoStop}%
\bibitem [{\citenamefont {Sanders}\ \emph {et~al.}(2016)\citenamefont
  {Sanders}, \citenamefont {Sanders}, \citenamefont {Strandjord}, \citenamefont
  {Qiu}, \citenamefont {Wu}, \citenamefont {Smiciklas}, \citenamefont {Mead},
  \citenamefont {Mosor}, \citenamefont {Arrizon}, \citenamefont {Ho},\ and\
  \citenamefont {Salit}}]{Sanders2016}%
  \BibitemOpen
  \bibfield  {author} {\bibinfo {author} {\bibfnamefont {G.~A.}\ \bibnamefont
  {Sanders}}, \bibinfo {author} {\bibfnamefont {S.~J.}\ \bibnamefont
  {Sanders}}, \bibinfo {author} {\bibfnamefont {L.~K.}\ \bibnamefont
  {Strandjord}}, \bibinfo {author} {\bibfnamefont {T.}~\bibnamefont {Qiu}},
  \bibinfo {author} {\bibfnamefont {J.}~\bibnamefont {Wu}}, \bibinfo {author}
  {\bibfnamefont {M.}~\bibnamefont {Smiciklas}}, \bibinfo {author}
  {\bibfnamefont {D.}~\bibnamefont {Mead}}, \bibinfo {author} {\bibfnamefont
  {S.}~\bibnamefont {Mosor}}, \bibinfo {author} {\bibfnamefont
  {A.}~\bibnamefont {Arrizon}}, \bibinfo {author} {\bibfnamefont
  {W.}~\bibnamefont {Ho}},\ and\ \bibinfo {author} {\bibfnamefont
  {M.}~\bibnamefont {Salit}},\ }in\ \href@noop {} {\emph {\bibinfo {booktitle}
  {Fiber Optic Sensors and Applications XIII}}},\ Vol.\ \bibinfo {volume}
  {9852},\ \bibinfo {editor} {edited by\ \bibinfo {editor} {\bibfnamefont
  {E.}~\bibnamefont {Udd}}, \bibinfo {editor} {\bibfnamefont {G.}~\bibnamefont
  {Pickrell}},\ and\ \bibinfo {editor} {\bibfnamefont {H.~H.}\ \bibnamefont
  {Du}}},\ \bibinfo {organization} {International Society for Optics and
  Photonics}\ (\bibinfo  {publisher} {SPIE},\ \bibinfo {year} {2016})\ pp.\
  \bibinfo {pages} {37 -- 50}\BibitemShut {NoStop}%
\bibitem [{\citenamefont {Napoli}\ and\ \citenamefont
  {Ward}(2016)}]{Napoli2016}%
  \BibitemOpen
  \bibfield  {author} {\bibinfo {author} {\bibfnamefont {J.}~\bibnamefont
  {Napoli}}\ and\ \bibinfo {author} {\bibfnamefont {R.}~\bibnamefont {Ward}},\
  }in\ \href@noop {} {\emph {\bibinfo {booktitle} {2016 DGON Intertial Sensors
  and Systems (ISS)}}}\ (\bibinfo {year} {2016})\ pp.\ \bibinfo {pages}
  {1--19}\BibitemShut {NoStop}%
\bibitem [{\citenamefont {Paturel}\ \emph {et~al.}(2014)\citenamefont
  {Paturel}, \citenamefont {Honthaas}, \citenamefont {Lef{\`e}vre},\ and\
  \citenamefont {Napolitano}}]{Paturel2014}%
  \BibitemOpen
  \bibfield  {author} {\bibinfo {author} {\bibfnamefont {Y.}~\bibnamefont
  {Paturel}}, \bibinfo {author} {\bibfnamefont {J.}~\bibnamefont {Honthaas}},
  \bibinfo {author} {\bibfnamefont {H.}~\bibnamefont {Lef{\`e}vre}},\ and\
  \bibinfo {author} {\bibfnamefont {F.}~\bibnamefont {Napolitano}},\
  }\href@noop {} {\bibfield  {journal} {\bibinfo  {journal} {Gyroscopy and
  Navigat.}\ }\textbf {\bibinfo {volume} {5}},\ \bibinfo {pages} {1} (\bibinfo
  {year} {2014})}\BibitemShut {NoStop}%
\bibitem [{\citenamefont {Schmelzbach}\ \emph {et~al.}(2018)\citenamefont
  {Schmelzbach}, \citenamefont {Donner}, \citenamefont {Igel}, \citenamefont
  {Sollberger}, \citenamefont {Taufiqurrahman}, \citenamefont {Bernauer},
  \citenamefont {Häusler}, \citenamefont {Renterghem}, \citenamefont
  {Wassermann},\ and\ \citenamefont {Robertsson}}]{Schmelzbach2018}%
  \BibitemOpen
  \bibfield  {author} {\bibinfo {author} {\bibfnamefont {C.}~\bibnamefont
  {Schmelzbach}}, \bibinfo {author} {\bibfnamefont {S.}~\bibnamefont {Donner}},
  \bibinfo {author} {\bibfnamefont {H.}~\bibnamefont {Igel}}, \bibinfo {author}
  {\bibfnamefont {D.}~\bibnamefont {Sollberger}}, \bibinfo {author}
  {\bibfnamefont {T.}~\bibnamefont {Taufiqurrahman}}, \bibinfo {author}
  {\bibfnamefont {F.}~\bibnamefont {Bernauer}}, \bibinfo {author}
  {\bibfnamefont {M.}~\bibnamefont {Häusler}}, \bibinfo {author}
  {\bibfnamefont {C.~V.}\ \bibnamefont {Renterghem}}, \bibinfo {author}
  {\bibfnamefont {J.}~\bibnamefont {Wassermann}},\ and\ \bibinfo {author}
  {\bibfnamefont {J.}~\bibnamefont {Robertsson}},\ }\href@noop {} {\bibfield
  {journal} {\bibinfo  {journal} {GEOPHYSICS}\ }\textbf {\bibinfo {volume}
  {83}},\ \bibinfo {pages} {WC53} (\bibinfo {year} {2018})}\BibitemShut
  {NoStop}%
\bibitem [{\citenamefont {Kurzych}\ \emph {et~al.}(2018)\citenamefont
  {Kurzych}, \citenamefont {Jaroszewicz}, \citenamefont {Krajewski},
  \citenamefont {Sakowicz}, \citenamefont {Kowalski},\ and\ \citenamefont
  {Mar{\'c}}}]{Kurzych2018}%
  \BibitemOpen
  \bibfield  {author} {\bibinfo {author} {\bibfnamefont {A.}~\bibnamefont
  {Kurzych}}, \bibinfo {author} {\bibfnamefont {L.~R.}\ \bibnamefont
  {Jaroszewicz}}, \bibinfo {author} {\bibfnamefont {Z.}~\bibnamefont
  {Krajewski}}, \bibinfo {author} {\bibfnamefont {B.}~\bibnamefont {Sakowicz}},
  \bibinfo {author} {\bibfnamefont {J.~K.}\ \bibnamefont {Kowalski}},\ and\
  \bibinfo {author} {\bibfnamefont {P.}~\bibnamefont {Mar{\'c}}},\ }\href@noop
  {} {\bibfield  {journal} {\bibinfo  {journal} {J. Lightwave Technol.}\
  }\textbf {\bibinfo {volume} {36}},\ \bibinfo {pages} {879} (\bibinfo {year}
  {2018})}\BibitemShut {NoStop}%
\bibitem [{\citenamefont {Toldi}\ \emph {et~al.}(2017)\citenamefont {Toldi},
  \citenamefont {de~Toldi}, \citenamefont {Lef{\`e}vre}, \citenamefont
  {Guattari}, \citenamefont {Bigueur}, \citenamefont {Steib}, \citenamefont
  {Ponceau}, \citenamefont {Molucon}, \citenamefont {Ducloux}, \citenamefont
  {Wassermann},\ and\ \citenamefont {Schreiber}}]{Toldi2017}%
  \BibitemOpen
  \bibfield  {author} {\bibinfo {author} {\bibfnamefont {E.~d.}\ \bibnamefont
  {Toldi}}, \bibinfo {author} {\bibfnamefont {E.}~\bibnamefont {de~Toldi}},
  \bibinfo {author} {\bibfnamefont {H.}~\bibnamefont {Lef{\`e}vre}}, \bibinfo
  {author} {\bibfnamefont {F.}~\bibnamefont {Guattari}}, \bibinfo {author}
  {\bibfnamefont {A.}~\bibnamefont {Bigueur}}, \bibinfo {author} {\bibfnamefont
  {A.}~\bibnamefont {Steib}}, \bibinfo {author} {\bibfnamefont
  {D.}~\bibnamefont {Ponceau}}, \bibinfo {author} {\bibfnamefont
  {C.}~\bibnamefont {Molucon}}, \bibinfo {author} {\bibfnamefont
  {E.}~\bibnamefont {Ducloux}}, \bibinfo {author} {\bibfnamefont
  {J.}~\bibnamefont {Wassermann}},\ and\ \bibinfo {author} {\bibfnamefont
  {U.}~\bibnamefont {Schreiber}},\ }\href@noop {} {\bibinfo {title} {First
  steps for a giant {FOG}: Searching for the limits}} (\bibinfo {year}
  {2017})\BibitemShut {NoStop}%
\bibitem [{\citenamefont {Li}\ \emph {et~al.}(2019)\citenamefont {Li},
  \citenamefont {Cao}, \citenamefont {He}, \citenamefont {Wu}, \citenamefont
  {Chen}, \citenamefont {Peng},\ and\ \citenamefont {Li}}]{Li2019}%
  \BibitemOpen
  \bibfield  {author} {\bibinfo {author} {\bibfnamefont {Y.}~\bibnamefont
  {Li}}, \bibinfo {author} {\bibfnamefont {Y.}~\bibnamefont {Cao}}, \bibinfo
  {author} {\bibfnamefont {D.}~\bibnamefont {He}}, \bibinfo {author}
  {\bibfnamefont {Y.}~\bibnamefont {Wu}}, \bibinfo {author} {\bibfnamefont
  {F.}~\bibnamefont {Chen}}, \bibinfo {author} {\bibfnamefont {C.}~\bibnamefont
  {Peng}},\ and\ \bibinfo {author} {\bibfnamefont {Z.}~\bibnamefont {Li}},\
  }\href@noop {} {\bibfield  {journal} {\bibinfo  {journal} {Opt. Express}\
  }\textbf {\bibinfo {volume} {27}},\ \bibinfo {pages} {14121} (\bibinfo {year}
  {2019})}\BibitemShut {NoStop}%
\bibitem [{\citenamefont {Guattari}\ \emph {et~al.}(2016)\citenamefont
  {Guattari}, \citenamefont {Molu{\c c}on}, \citenamefont {Bigueur},
  \citenamefont {Ducloux}, \citenamefont {de~Toldi}, \citenamefont {Honthaas},\
  and\ \citenamefont {Lef{\`e}vre}}]{Guattari2016}%
  \BibitemOpen
  \bibfield  {author} {\bibinfo {author} {\bibfnamefont {F.}~\bibnamefont
  {Guattari}}, \bibinfo {author} {\bibfnamefont {C.}~\bibnamefont {Molu{\c
  c}on}}, \bibinfo {author} {\bibfnamefont {A.}~\bibnamefont {Bigueur}},
  \bibinfo {author} {\bibfnamefont {E.}~\bibnamefont {Ducloux}}, \bibinfo
  {author} {\bibfnamefont {E.}~\bibnamefont {de~Toldi}}, \bibinfo {author}
  {\bibfnamefont {J.}~\bibnamefont {Honthaas}},\ and\ \bibinfo {author}
  {\bibfnamefont {H.}~\bibnamefont {Lef{\`e}vre}},\ }in\ \href@noop {} {\emph
  {\bibinfo {booktitle} {2016 {DGON} Intertial Sensors and Systems ({ISS})}}}\
  (\bibinfo {year} {2016})\ pp.\ \bibinfo {pages} {1--13}\BibitemShut {NoStop}%
\bibitem [{\citenamefont {Logozinski\u{i}}(1981)}]{Logozinskii1981}%
  \BibitemOpen
  \bibfield  {author} {\bibinfo {author} {\bibfnamefont {V.~N.}\ \bibnamefont
  {Logozinski\u{i}}},\ }\href@noop {} {\bibfield  {journal} {\bibinfo
  {journal} {Sov. J. Quantum Electron.}\ }\textbf {\bibinfo {volume} {11}},\
  \bibinfo {pages} {536} (\bibinfo {year} {1981})}\BibitemShut {NoStop}%
\bibitem [{\citenamefont {Knudsen}\ and\ \citenamefont
  {Bl{\o}tekj{\ae}r}(1995)}]{Knudsen1995}%
  \BibitemOpen
  \bibfield  {author} {\bibinfo {author} {\bibfnamefont {S.}~\bibnamefont
  {Knudsen}}\ and\ \bibinfo {author} {\bibfnamefont {K.}~\bibnamefont
  {Bl{\o}tekj{\ae}r}},\ }\href@noop {} {\bibfield  {journal} {\bibinfo
  {journal} {Opt. Lett.}\ }\textbf {\bibinfo {volume} {20}},\ \bibinfo {pages}
  {1432} (\bibinfo {year} {1995})}\BibitemShut {NoStop}%
\bibitem [{\citenamefont {Moeller}\ and\ \citenamefont
  {Burns}(1996)}]{Moeller1996}%
  \BibitemOpen
  \bibfield  {author} {\bibinfo {author} {\bibfnamefont {R.~P.}\ \bibnamefont
  {Moeller}}\ and\ \bibinfo {author} {\bibfnamefont {W.~K.}\ \bibnamefont
  {Burns}},\ }\href@noop {} {\bibfield  {journal} {\bibinfo  {journal} {Opt.
  Lett.}\ }\textbf {\bibinfo {volume} {21}},\ \bibinfo {pages} {171} (\bibinfo
  {year} {1996})}\BibitemShut {NoStop}%
\bibitem [{\citenamefont {Morris}\ \emph {et~al.}(2022)\citenamefont {Morris},
  \citenamefont {Zawada}, \citenamefont {Garcia}, \citenamefont {Wheeler},\
  and\ \citenamefont {Digonnet}}]{Morris2022}%
  \BibitemOpen
  \bibfield  {author} {\bibinfo {author} {\bibfnamefont {T.~A.}\ \bibnamefont
  {Morris}}, \bibinfo {author} {\bibfnamefont {A.~N.}\ \bibnamefont {Zawada}},
  \bibinfo {author} {\bibfnamefont {D.}~\bibnamefont {Garcia}}, \bibinfo
  {author} {\bibfnamefont {J.~M.}\ \bibnamefont {Wheeler}},\ and\ \bibinfo
  {author} {\bibfnamefont {M.~J.~F.}\ \bibnamefont {Digonnet}},\ }\href@noop {}
  {\bibfield  {journal} {\bibinfo  {journal} {IEEE Sens. J.}\ }\textbf
  {\bibinfo {volume} {22}},\ \bibinfo {pages} {2205} (\bibinfo {year}
  {2022})}\BibitemShut {NoStop}%
\bibitem [{\citenamefont {Burns}\ \emph {et~al.}(1990)\citenamefont {Burns},
  \citenamefont {Moeller},\ and\ \citenamefont {Dandridge}}]{Burns1990}%
  \BibitemOpen
  \bibfield  {author} {\bibinfo {author} {\bibfnamefont {W.~K.}\ \bibnamefont
  {Burns}}, \bibinfo {author} {\bibfnamefont {R.~P.}\ \bibnamefont {Moeller}},\
  and\ \bibinfo {author} {\bibfnamefont {A.}~\bibnamefont {Dandridge}},\
  }\href@noop {} {\bibfield  {journal} {\bibinfo  {journal} {IEEE Photonics
  Technology Letters}\ }\textbf {\bibinfo {volume} {2}},\ \bibinfo {pages}
  {606} (\bibinfo {year} {1990})}\BibitemShut {NoStop}%
\bibitem [{\citenamefont {Burns}\ and\ \citenamefont
  {Moeller}(1996)}]{Burns1996}%
  \BibitemOpen
  \bibfield  {author} {\bibinfo {author} {\bibfnamefont {W.~K.}\ \bibnamefont
  {Burns}}\ and\ \bibinfo {author} {\bibfnamefont {R.~P.}\ \bibnamefont
  {Moeller}},\ }in\ \href@noop {} {\emph {\bibinfo {booktitle} {Fiber Optic
  Gyros: 20th Anniversary Conference}}},\ Vol.\ \bibinfo {volume} {2837},\
  \bibinfo {editor} {edited by\ \bibinfo {editor} {\bibfnamefont
  {E.}~\bibnamefont {Udd}}, \bibinfo {editor} {\bibfnamefont {H.~C.}\
  \bibnamefont {Lefevre}},\ and\ \bibinfo {editor} {\bibfnamefont
  {K.}~\bibnamefont {Hotate}}},\ \bibinfo {organization} {International Society
  for Optics and Photonics}\ (\bibinfo  {publisher} {SPIE},\ \bibinfo {year}
  {1996})\ pp.\ \bibinfo {pages} {381 -- 387}\BibitemShut {NoStop}%
\bibitem [{\citenamefont {Shin}\ \emph {et~al.}(2010)\citenamefont {Shin},
  \citenamefont {Sharma}, \citenamefont {Tu}, \citenamefont {Jung},\ and\
  \citenamefont {Boppart}}]{Shin2010}%
  \BibitemOpen
  \bibfield  {author} {\bibinfo {author} {\bibfnamefont {S.}~\bibnamefont
  {Shin}}, \bibinfo {author} {\bibfnamefont {U.}~\bibnamefont {Sharma}},
  \bibinfo {author} {\bibfnamefont {H.}~\bibnamefont {Tu}}, \bibinfo {author}
  {\bibfnamefont {W.}~\bibnamefont {Jung}},\ and\ \bibinfo {author}
  {\bibfnamefont {S.~A.}\ \bibnamefont {Boppart}},\ }\href@noop {} {\bibfield
  {journal} {\bibinfo  {journal} {IEEE Photonics Technol. Lett.}\ }\textbf
  {\bibinfo {volume} {22}},\ \bibinfo {pages} {1057} (\bibinfo {year}
  {2010})}\BibitemShut {NoStop}%
\bibitem [{\citenamefont {Blake}\ and\ \citenamefont {Kim}(1994)}]{Blake1994}%
  \BibitemOpen
  \bibfield  {author} {\bibinfo {author} {\bibfnamefont {J.}~\bibnamefont
  {Blake}}\ and\ \bibinfo {author} {\bibfnamefont {I.~S.}\ \bibnamefont
  {Kim}},\ }\href@noop {} {\bibfield  {journal} {\bibinfo  {journal} {Opt.
  Lett.}\ }\textbf {\bibinfo {volume} {19}},\ \bibinfo {pages} {1648} (\bibinfo
  {year} {1994})}\BibitemShut {NoStop}%
\bibitem [{\citenamefont {Yurek}\ \emph {et~al.}(1986)\citenamefont {Yurek},
  \citenamefont {Taylor}, \citenamefont {Goldberg}, \citenamefont {Weller},\
  and\ \citenamefont {Dandridge}}]{Yurek1986}%
  \BibitemOpen
  \bibfield  {author} {\bibinfo {author} {\bibfnamefont {A.~M.}\ \bibnamefont
  {Yurek}}, \bibinfo {author} {\bibfnamefont {H.~F.}\ \bibnamefont {Taylor}},
  \bibinfo {author} {\bibfnamefont {L.}~\bibnamefont {Goldberg}}, \bibinfo
  {author} {\bibfnamefont {J.~F.}\ \bibnamefont {Weller}},\ and\ \bibinfo
  {author} {\bibfnamefont {A.}~\bibnamefont {Dandridge}},\ }\href@noop {}
  {\bibfield  {journal} {\bibinfo  {journal} {IEEE Journal of Quantum
  Electronics}\ }\textbf {\bibinfo {volume} {22}},\ \bibinfo {pages} {522}
  (\bibinfo {year} {1986})}\BibitemShut {NoStop}%
\bibitem [{\citenamefont {Moeller}\ and\ \citenamefont
  {Burns}(1991)}]{Moeller1991}%
  \BibitemOpen
  \bibfield  {author} {\bibinfo {author} {\bibfnamefont {R.~P.}\ \bibnamefont
  {Moeller}}\ and\ \bibinfo {author} {\bibfnamefont {W.~K.}\ \bibnamefont
  {Burns}},\ }\href@noop {} {\bibfield  {journal} {\bibinfo  {journal} {Opt.
  Lett.}\ }\textbf {\bibinfo {volume} {16}},\ \bibinfo {pages} {1902} (\bibinfo
  {year} {1991})}\BibitemShut {NoStop}%
\bibitem [{\citenamefont {Guattari}\ \emph {et~al.}(2014)\citenamefont
  {Guattari}, \citenamefont {Chouvin}, \citenamefont {Molu{\c c}on},\ and\
  \citenamefont {Lef{\`e}vre}}]{Guattari2014}%
  \BibitemOpen
  \bibfield  {author} {\bibinfo {author} {\bibfnamefont {F.}~\bibnamefont
  {Guattari}}, \bibinfo {author} {\bibfnamefont {S.}~\bibnamefont {Chouvin}},
  \bibinfo {author} {\bibfnamefont {C.}~\bibnamefont {Molu{\c c}on}},\ and\
  \bibinfo {author} {\bibfnamefont {H.}~\bibnamefont {Lef{\`e}vre}},\ }in\
  \href@noop {} {\emph {\bibinfo {booktitle} {2014 {DGON} Inertial Sensors and
  Syst. ({ISS})}}}\ (\bibinfo {year} {2014})\ pp.\ \bibinfo {pages}
  {1--14}\BibitemShut {NoStop}%
\bibitem [{\citenamefont {Polynkin}\ \emph {et~al.}(2000)\citenamefont
  {Polynkin}, \citenamefont {de~Arruda},\ and\ \citenamefont
  {Blake}}]{Polynkin2000}%
  \BibitemOpen
  \bibfield  {author} {\bibinfo {author} {\bibfnamefont {P.}~\bibnamefont
  {Polynkin}}, \bibinfo {author} {\bibfnamefont {J.}~\bibnamefont
  {de~Arruda}},\ and\ \bibinfo {author} {\bibfnamefont {J.}~\bibnamefont
  {Blake}},\ }\href@noop {} {\bibfield  {journal} {\bibinfo  {journal} {Opt.
  Lett.}\ }\textbf {\bibinfo {volume} {25}},\ \bibinfo {pages} {147} (\bibinfo
  {year} {2000})}\BibitemShut {NoStop}%
\bibitem [{\citenamefont {Rabelo}\ \emph {et~al.}(2000)\citenamefont {Rabelo},
  \citenamefont {de~Carvalho},\ and\ \citenamefont {Blake}}]{Rabelo2000}%
  \BibitemOpen
  \bibfield  {author} {\bibinfo {author} {\bibfnamefont {R.~C.}\ \bibnamefont
  {Rabelo}}, \bibinfo {author} {\bibfnamefont {R.~T.}\ \bibnamefont
  {de~Carvalho}},\ and\ \bibinfo {author} {\bibfnamefont {J.}~\bibnamefont
  {Blake}},\ }\href@noop {} {\bibfield  {journal} {\bibinfo  {journal} {J.
  Lightwave Technol.}\ }\textbf {\bibinfo {volume} {18}},\ \bibinfo {pages}
  {2146} (\bibinfo {year} {2000})}\BibitemShut {NoStop}%
\bibitem [{\citenamefont {Honthaas}\ \emph {et~al.}(2014)\citenamefont
  {Honthaas}, \citenamefont {Bonnefois}, \citenamefont {Ducloux},\ and\
  \citenamefont {Lefèvre}}]{Honthaas2014}%
  \BibitemOpen
  \bibfield  {author} {\bibinfo {author} {\bibfnamefont {J.}~\bibnamefont
  {Honthaas}}, \bibinfo {author} {\bibfnamefont {J.-J.}\ \bibnamefont
  {Bonnefois}}, \bibinfo {author} {\bibfnamefont {E.}~\bibnamefont {Ducloux}},\
  and\ \bibinfo {author} {\bibfnamefont {H.}~\bibnamefont {Lefèvre}},\ }in\
  \href@noop {} {\emph {\bibinfo {booktitle} {23rd International Conference on
  Optical Fibre Sensors}}},\ Vol.\ \bibinfo {volume} {9157},\ \bibinfo {editor}
  {edited by\ \bibinfo {editor} {\bibfnamefont {J.~M.}\ \bibnamefont
  {López-Higuera}}, \bibinfo {editor} {\bibfnamefont {J.~D.~C.}\ \bibnamefont
  {Jones}}, \bibinfo {editor} {\bibfnamefont {M.}~\bibnamefont {López-Amo}},\
  and\ \bibinfo {editor} {\bibfnamefont {J.~L.}\ \bibnamefont {Santos}}},\
  \bibinfo {organization} {International Society for Optics and Photonics}\
  (\bibinfo  {publisher} {SPIE},\ \bibinfo {year} {2014})\ pp.\ \bibinfo
  {pages} {329 -- 332}\BibitemShut {NoStop}%
\bibitem [{\citenamefont {He}\ \emph {et~al.}(2020)\citenamefont {He},
  \citenamefont {Cao}, \citenamefont {Zhou}, \citenamefont {Peng},\ and\
  \citenamefont {Li}}]{He2020}%
  \BibitemOpen
  \bibfield  {author} {\bibinfo {author} {\bibfnamefont {D.}~\bibnamefont
  {He}}, \bibinfo {author} {\bibfnamefont {Y.}~\bibnamefont {Cao}}, \bibinfo
  {author} {\bibfnamefont {T.}~\bibnamefont {Zhou}}, \bibinfo {author}
  {\bibfnamefont {C.}~\bibnamefont {Peng}},\ and\ \bibinfo {author}
  {\bibfnamefont {Z.}~\bibnamefont {Li}},\ }\href@noop {} {\bibfield  {journal}
  {\bibinfo  {journal} {Opt. Express}\ }\textbf {\bibinfo {volume} {28}},\
  \bibinfo {pages} {34717} (\bibinfo {year} {2020})}\BibitemShut {NoStop}%
\bibitem [{\citenamefont {Suo}\ \emph {et~al.}(2021)\citenamefont {Suo},
  \citenamefont {Yu}, \citenamefont {Yang}, \citenamefont {Feng}, \citenamefont
  {Xie}, \citenamefont {Chang}, \citenamefont {He}, \citenamefont {Tang},\ and\
  \citenamefont {Xiang}}]{Suo2021}%
  \BibitemOpen
  \bibfield  {author} {\bibinfo {author} {\bibfnamefont {X.}~\bibnamefont
  {Suo}}, \bibinfo {author} {\bibfnamefont {H.}~\bibnamefont {Yu}}, \bibinfo
  {author} {\bibfnamefont {Y.}~\bibnamefont {Yang}}, \bibinfo {author}
  {\bibfnamefont {W.}~\bibnamefont {Feng}}, \bibinfo {author} {\bibfnamefont
  {P.}~\bibnamefont {Xie}}, \bibinfo {author} {\bibfnamefont {Y.}~\bibnamefont
  {Chang}}, \bibinfo {author} {\bibfnamefont {Y.}~\bibnamefont {He}}, \bibinfo
  {author} {\bibfnamefont {M.}~\bibnamefont {Tang}},\ and\ \bibinfo {author}
  {\bibfnamefont {Z.}~\bibnamefont {Xiang}},\ }\href@noop {} {\bibfield
  {journal} {\bibinfo  {journal} {Appl. Opt.}\ }\textbf {\bibinfo {volume}
  {60}},\ \bibinfo {pages} {3103} (\bibinfo {year} {2021})}\BibitemShut
  {NoStop}%
\bibitem [{MM2()}]{MM2022}%
  \BibitemOpen
  \href@noop {} {}\bibinfo {note} {The detail of theoretical derivation will be
  published elsewhere.}\BibitemShut {Stop}%
\bibitem [{\citenamefont {Moeller}\ \emph {et~al.}(1989)\citenamefont
  {Moeller}, \citenamefont {Burns},\ and\ \citenamefont {Frigo}}]{Moeller1989}%
  \BibitemOpen
  \bibfield  {author} {\bibinfo {author} {\bibfnamefont {R.~P.}\ \bibnamefont
  {Moeller}}, \bibinfo {author} {\bibfnamefont {W.~K.}\ \bibnamefont {Burns}},\
  and\ \bibinfo {author} {\bibfnamefont {N.~J.}\ \bibnamefont {Frigo}},\
  }\href@noop {} {\bibfield  {journal} {\bibinfo  {journal} {J. Lightwave
  Technol.}\ }\textbf {\bibinfo {volume} {7}},\ \bibinfo {pages} {262}
  (\bibinfo {year} {1989})}\BibitemShut {NoStop}%
\bibitem [{\citenamefont {Scott}\ \emph {et~al.}(2001)\citenamefont {Scott},
  \citenamefont {Langrock},\ and\ \citenamefont {Kolner}}]{Scott2001}%
  \BibitemOpen
  \bibfield  {author} {\bibinfo {author} {\bibfnamefont {R.~P.}\ \bibnamefont
  {Scott}}, \bibinfo {author} {\bibfnamefont {C.}~\bibnamefont {Langrock}},\
  and\ \bibinfo {author} {\bibfnamefont {B.~H.}\ \bibnamefont {Kolner}},\
  }\href@noop {} {\bibfield  {journal} {\bibinfo  {journal} {IEEE J. Sel. Top.
  Quantum Electron.}\ }\textbf {\bibinfo {volume} {7}},\ \bibinfo {pages} {641}
  (\bibinfo {year} {2001})}\BibitemShut {NoStop}%
\bibitem [{\citenamefont {B\"{o}hm}\ \emph {et~al.}(1983)\citenamefont
  {B\"{o}hm}, \citenamefont {Marten}, \citenamefont {Weidel},\ and\
  \citenamefont {Petermann}}]{Bohm1983}%
  \BibitemOpen
  \bibfield  {author} {\bibinfo {author} {\bibfnamefont {K.}~\bibnamefont
  {B\"{o}hm}}, \bibinfo {author} {\bibfnamefont {P.}~\bibnamefont {Marten}},
  \bibinfo {author} {\bibfnamefont {E.}~\bibnamefont {Weidel}},\ and\ \bibinfo
  {author} {\bibfnamefont {K.}~\bibnamefont {Petermann}},\ }\href@noop {}
  {\bibfield  {journal} {\bibinfo  {journal} {Electronics Letters}\ }\textbf
  {\bibinfo {volume} {19}},\ \bibinfo {pages} {997} (\bibinfo {year}
  {1983})}\BibitemShut {NoStop}%
\bibitem [{\citenamefont {Chardon}\ and\ \citenamefont
  {Huard}(1983)}]{Chardon1983}%
  \BibitemOpen
  \bibfield  {author} {\bibinfo {author} {\bibfnamefont {D.}~\bibnamefont
  {Chardon}}\ and\ \bibinfo {author} {\bibfnamefont {S.~J.}\ \bibnamefont
  {Huard}},\ }\href@noop {} {\bibfield  {journal} {\bibinfo  {journal}
  {Canadian Journal of Physics}\ }\textbf {\bibinfo {volume} {61}},\ \bibinfo
  {pages} {1334} (\bibinfo {year} {1983})}\BibitemShut {NoStop}%
\bibitem [{IEE(1998)}]{IEEE1998}%
  \BibitemOpen
  \href@noop {} {\bibinfo {title} {\uppercase{IEEE} standard specification
  format guide and test procedure for single-axis interferometric fiber optic
  gyros}},\ \bibinfo {howpublished} {{\it IEEE Std 952-1997}} (\bibinfo {year}
  {1998})\BibitemShut {NoStop}%
\end{thebibliography}%

\end{document}